\documentclass[prl,twocolumn,showpacs,preprintnumbers,amsmath,amssymb,superscriptaddress,longbibliography]{revtex4-1}

\usepackage{amsmath}    % need for subequations
\usepackage{graphicx}   % need for figures
\usepackage{verbatim}   % useful for program listings
\usepackage{color}      % use if color is used in text
\usepackage{subfigure}  % use for side-by-side figures
\usepackage{hyperref}   % use for hypertext links, including those to external documents and URLs
\usepackage{multirow}
\usepackage{float}
\usepackage{braket}
\usepackage{siunitx}
\usepackage{bbm}
\usepackage{mathbbol}
\usepackage{slashed}

\newcommand{\nn}{\nonumber}

\begin{document}

\title{Neural-Network Approach to Dissipative Quantum Many-Body Dynamics}
\author{Michael J. Hartmann}
\affiliation{Institute of Photonics and Quantum Sciences, Heriot-Watt University Edinburgh EH14 4AS, United Kingdom}
%\email{present address: Google Inc., Erika-Mann-Str. 33, 80636 M\"unchen, Germany}
\affiliation{Google Research, Erika-Mann-Str. 33, 80636 M\"unchen, Germany}
\author{Giuseppe Carleo}
\affiliation{Center for Computational Quantum Physics, Flatiron Institute,
162 5th Avenue, New York, NY 10010, USA}
\date{\today}

\begin{abstract}
In experimentally realistic situations, quantum systems are never perfectly isolated and the coupling to their environment needs to be taken into account. Often, the effect of the environment can be well approximated by a Markovian master equation. However, solving this master equation for quantum many-body systems, becomes exceedingly hard due to the high dimension of the Hilbert space. Here we present an approach to the effective simulation of the dynamics of open quantum many-body systems based on machine learning techniques. We represent the mixed many-body quantum states with neural networks in the form of restricted Boltzmann machines and derive a variational Monte-Carlo algorithm for their time evolution and stationary states. We document the accuracy of the approach with numerical examples for a dissipative spin lattice system.
\end{abstract}

\maketitle

The description of interacting quantum many-body systems presents a formidable challenge for theoretical and numerical approaches. A pure many-body quantum state is described by the wave-function, whose complexity grows exponentially with the number of constituents. This challenge is even more pronounced for mixed quantum states, where the fundamental object describing all physical properties is the density matrix, whose degrees of freedom scale quadratically with the dimension of the Hilbert space \cite{breuer_theory_2007}. Yet the description of experiments under realistic conditions requires modeling in terms of density matrices as the systems of interest are never perfectly isolated from their environment. The huge number of degrees of freedom of pure and mixed states however renders an exact description of large systems in general infeasible (see \cite{Prosen11,Prosen14} for exceptions with exact solutions), even if one resorts to numerical approaches. 

To meet this quantum complexity challenge, several approximate approaches have been developed. Tensor Networks and the Density Matrix Renormalization Group \cite{schollwock_density-matrix_2011,verstraete_matrix_2008} become efficient descriptions whenever the amount of entanglement contained in the modeled states is restricted. Despite substantial effort \cite{kshetrimayum_simple_2017}, these methods however still suffer from limitations in systems with more than one lattice dimension.  For two-dimensional systems, novel real-space renormalization-based approaches are among the most promising existing tools \cite{finazzi_corner-space_2015}, although their application to large systems is at present at the forefront of research activity \cite{rota_quantum_2018}.  Stochastic many-body techniques, such as Quantum Monte Carlo (QMC) methods \cite{ceperley_quantum_1986,foulkes_quantum_2001} rely on sampling a number of physically relevant configurations or perform an efficient compression of the quantum state. However, QMC approaches are effective only for a restricted number of open quantum systems and regimes \cite{yan_interacting_2018,nagy_driven-dissipative_2018}, and a severe sign problem typically emerges in the simulation of dissipative dynamics.  

Recently, machine-learning inspired approaches and parameterizations of wave-functions in terms of neural networks have been introduced\cite{carleo_solving_2017}. This variational representation, dubbed neural-network quantum states (NQS), has been used to study both system at equilibrium \cite{carleo_solving_2017,deng_machine_2017,glasser_neural-network_2018,kaubruegger_chiral_2018,choo_symmetries_2018}, and out-of-equilibrium, in the context of unitary dynamics of pure states \cite{schmitt_quantum_2018,czischek_quenches_2018,jonsson_neural-network_2018}. The connection between NQS and tensor network representations has also been explored \cite{glasser_neural-network_2018,chen_equivalence_2018,pastori_generalized_2018}. While in the past years there has been several methodological developments to study open quantum systems using Tensor Network representations \cite{verstraete_matrix_2004,zwolak_mixed-state_2004,orus_infinite_2008,cui_variational_2015,mascarenhas_matrix-product-operator_2015,werner_positive_2016,gangat_steady_2017,jaschke_one-dimensional_2018}, the description of mixed states with NQS has been so-far explored for data-driven tomographic purposes \cite{torlai_latent_2018,carrasquilla_reconstructing_2018,banchi_modelling_2018}. 

For modeling quantum experiments, particularly for open many-body systems \cite{Diehl:2008yq, Barreiro:2011qy, Fitzpatrick17, Collodo18, Ma18}, there is a strong need for efficient and accurate approaches, especially in more than one lattice dimension, where Tensor Networks face difficulties. To this end, it is instrumental to develop a flexible, and scalable numerical approach to study mixed state dynamics or stationary states of dissipative dynamics. Central to this goal is the ability to use variational density-matrix states not facing the entanglement problem, and flexible enough to describe correlations and many-body effects beyond mean-field \cite{marino_quantum_2016,casteels_gutzwiller_2018}, and cluster approaches \cite{jin_cluster_2016,biella_linked_2018}.  

Here we present a machine learning approach to the simulation of dissipative quantum dynamics and its stationary states. Our approach uses a neural network parameterization for the quantum density matrix \cite{torlai_latent_2018} and a stochastic learning method to approximate its dynamics in a time-dependent Variational Monte Carlo approach \cite{carleo_localization_2012}. Our approach is suitable to model non-unitary dynamics of quantum systems with many degrees of freedom in a variety of settings. These include numerical characterizations of near term quantum computers where decoherence processes due to their imperfections are taken into account \cite{Preskill18}. A second field of applications is the investigation of stationary state quantum phases and phase transitions, which have attracted increasing interest in recent years \cite{kessler_dissipative_2012, Diehl:2008yq, Barreiro:2011qy, Fitzpatrick17, Collodo18, Ma18, hartmann_quantum_2016}.

{\bf Problem and parameterization:}
Our aim is to solve the quantum master equation of Lindblad form,
\begin{equation}\label{eq:master}
	\dot{\rho} = - i [H,\rho] + \sum_j \frac{\gamma_j}{2} \left(2 c_j \rho c_j^\dag - c_j^\dag c_j \rho - \rho c_j^\dag c_j \right)
\end{equation}
where $\rho$ is the density matrix of the system, $H$ its Hamiltonian, and the $\gamma_j$ and $c_j$ the dissipation rates and jump operators of its dissipation. The index $j$ runs over all dissipation channels. For a large class of models, there is however only one dissipation channel per lattice site and we will restrict our treatment to this case, where $j$ thus labels the lattice sites.
As an example, we will consider a dissipative and anisotropic Heisenberg model for a lattice of $N$ spin-1/2 degrees of freedom that has attracted significant interest recently \cite{jin_cluster_2016}.

\begin{figure}
    \centering  \includegraphics[width=0.9\columnwidth]{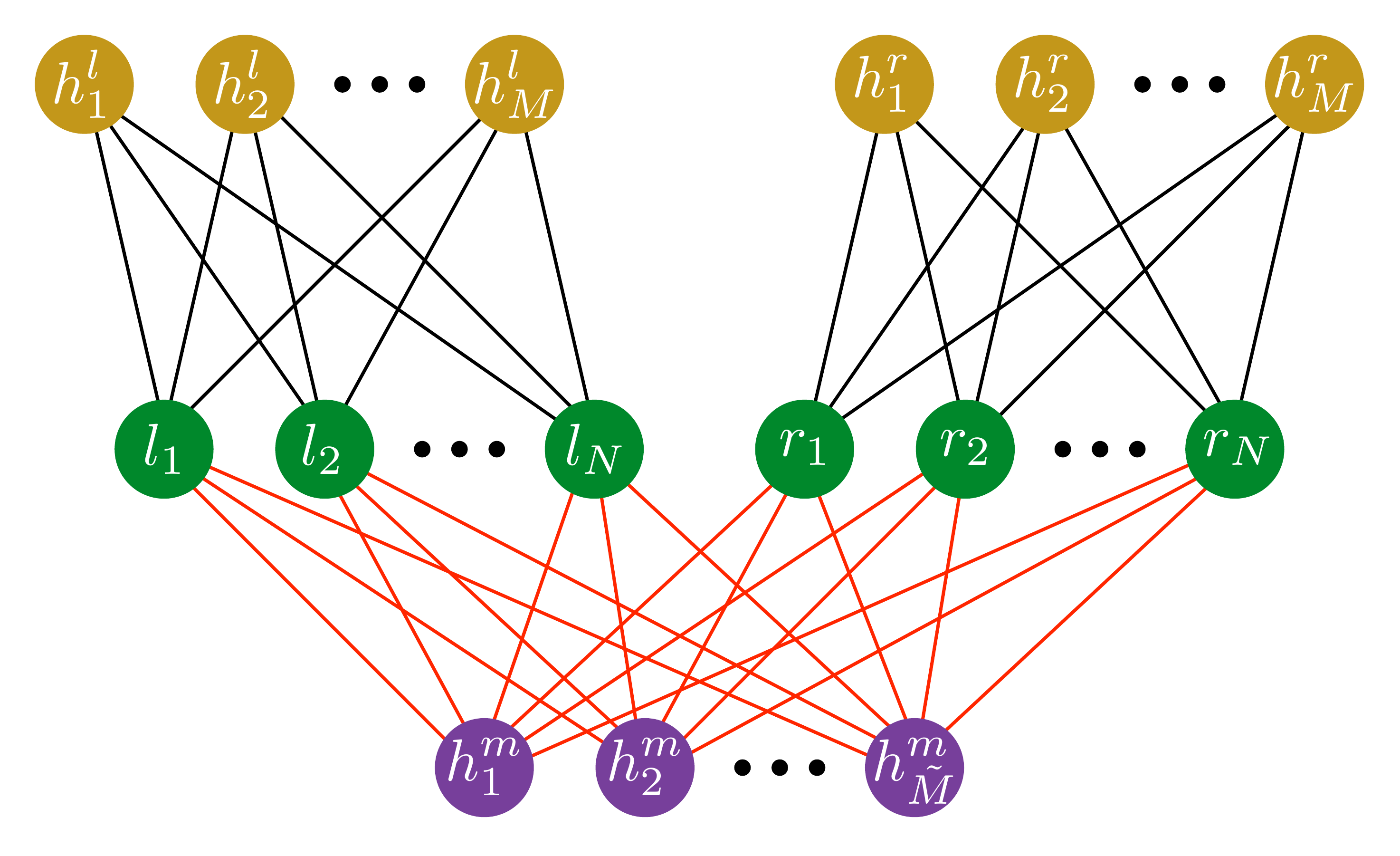}
	\caption{Sketch of the employed neural network. Visible layer in green and hidden layers in light brown and purple. There is a hidden layer for the row indices $l_i$ and the column indices $r_i$ of $\rho$ ($i= 1,2,\dots, N$) with hidden neurons $h_j^l$ and $h_j^r$ ($j=1,\dots,M$). A further hidden layer with neurons $h^m_k$ ($k=1,\dots,\tilde{M}$) is responsible for the mixing, c.f. \cite{torlai_latent_2018}.}
	\label{fig:network}
\end{figure}

To find an efficient and accurate approximation to the dynamics of Eq. (\ref{eq:master}), we leverage the idea that artificial neural networks can be used to provide compact representations of quantum states  \cite{carleo_solving_2017}. Specifically, we use a parametrization of the density matrix in terms of complex-valued Restricted Boltzmann Machines (RBM), similar to the one introduced in Ref. \cite{torlai_latent_2018}. Fig. \ref{fig:network} shows a sketch of the specific neural network architecture used in this work. 
It most prominently features three sets of hidden units, $h^{(l)}$, $h^{(r)}$ and $h^{(m)}$, whose role is to mediate correlations among, respectively, column degrees of freedom of the density matrix, row degrees of freedom, and mixed correlations between the two. Because of the bipartite structure of the RBM interactions, the hidden units can be integrated out exactly, resulting in a parametrization that guarantees a Hermitian and positive semi definite density matrix \cite{torlai_latent_2018},
\begin{align} \label{eq:parameter-torlai-2}
	\rho_{\vec{l},\vec{r}} & = \exp\left[\sum_{j=1}^N (a_j l_j + a_j^* r_j) \right] \times \prod_{k = 1}^M \mathcal{X}_k \times \prod_{p = 1}^{\tilde{M}} \mathcal{Y}_p\\
	\mathcal{X}_k & = \cosh\left(b_k + \sum_{j=1}^N W_{k,j} l_j\right) \cosh\left(b_k^* + \sum_{j=1}^N W_{k,j}^* r_j\right) \nn \\
	\mathcal{Y}_p & = \cosh\left(c_p + c_p^* + \sum_{j=1}^N (U_{p,j} l_j + U_{p,j}^* r_j)\right), \nn 
\end{align}
where the vector indices $\vec{l} = (l_1,l_2,\dots)$ and $\vec{r} = (r_1,r_2,\dots)$ contain the left (right) indices $l_j(r_j)$ for all lattice sites $j$, and the variational parameters are the complex-valued weights $W_{k,j}$, $U_{p,j}$ and biases $a_j$, $b_k$ and $c_p$. Analogously to the pure state case, increasing the number of hidden units, $M$ and $\tilde{M}$, guarantees more expressive representations of the density matrix. Given the RBM parametrization of the density matrix, it remains to be determined how to find an approximate solution of the Lindblad master equation. The approximation of the dynamics generated by Eq. (\ref{eq:master}) can be recast as a variational optimization problem, that can be approached via a suitable extension of the stochastic reconfiguration method \cite{sorella_weak_2007} and the time-dependent Variational Monte Carlo \cite{carleo_localization_2012} to the dissipative case.

{\bf Stochastic Reconfiguration for Liouvillians:}
It is convenient to write the density matrix $\rho$ as a vector $\vec{\rho}$ such that the right hand side of Eq. (\ref{eq:master}) can be expressed as the action of a linear operator on $\vec{\rho}$, i.e. $
    \partial_t \vec{\rho} = \mathcal{L} \vec{\rho}$, where $\mathcal{L}$ is the Liouvillian super-operator which, in contrast to the Hamiltonian $H$, is not Hermitian.

According to Eq. (\ref{eq:parameter-torlai-2}), the density matrix $\vec{\rho}$ is parameterized by a set of $(N+1) (M + \tilde{M}) + N$ complex variational parameters. In the following, we use the abbreviate notation $\vec{\alpha}$, to indicate the ensemble of these variational parameters. Most notably, the real vector $\vec{\alpha}$ contains both imaginary and real parts of the variational parameters, that are treated independently. The time derivative of the variational $\rho$ can in turn be expressed in terms of the time derivative of the variational parameters as,
\begin{equation}
	\partial_t \vec{\rho} = \sum_k \dot{\alpha}_k O_k \vec{\rho}
\end{equation}
where $O_k$ denote diagonal matrices whose non-zero matrix elements read,
$[O_k]_{\vec{l},\vec{r};\vec{l},\vec{r}} = \partial \ln ( \rho_{\vec{l},\vec{r}})/ (\partial \alpha_k)$.
%\begin{equation} \label{eq:log-devs}
%	[O_k]_{\vec{l},\vec{r};\vec{l},\vec{r}} = \frac{\partial}{\partial \alpha_k} \ln ( \rho_{\vec{l},\vec{r}})
%\end{equation}
To get the best approximation to the dynamics of the density matrix, our goal is to find a closed equation of motion for the variational parameters, namely the time-dependence 
$\alpha(t)$. To this end, at each instant in time we consider the difference between the exact Lindblad infinitesimal time evolution and the approximate variational evolution,
\begin{equation} \label{eq:twonorm}
	\delta = \left|\left| \sum_k \dot{\alpha}_k O_k \vec{\rho} - \mathcal{L} \vec{\rho}\right|\right|_2^2,
\end{equation}
where the time-derivatives of the variational parameters, $\dot{\alpha}_k$, are to be determined.  
Minimization of $\delta$ with respect to $\dot{\alpha}_k$ leads to the system of equations
\begin{equation} \label{eq:para_derivative}
	\sum_{p} S_{k,p} \, \dot\alpha_{p}= f_k
\end{equation}
where
\begin{eqnarray}
	S_{k,p} & = \vec{\rho}^\dag O_k^\dag O_p \vec{\rho}  + \vec{\rho}^\dag O_p^\dag O_k \vec{\rho}  \label{eq:eom-var-1} \\ 
	f_k & = \vec{\rho}^\dag O_k^\dag \mathcal{L} \vec{\rho} + \vec{\rho}^\dag \mathcal{L}^\dag O_k \vec{\rho}  \label{eq:eom-var-2}
\end{eqnarray}
and it is easy to show that the solutions of Eqs. (\ref{eq:para_derivative}) are indeed local minima of $\delta$, see Supplemental Material. Alternatively to the 2-norm in Eq. (\ref{eq:twonorm}), one can also use the Fubini-Study norm, see Supplemental Material.  Eq. (\ref{eq:para_derivative}) can be written as a first order differential equation,
\begin{equation} \label{eq:para_derivative_2}
	\partial_t \vec{\alpha} = \mathcal{S}^{-1} \, \vec{f},
\end{equation}
where $S_{k,p}$ are the matrix elements of the matrix $\mathcal{S}$ and $f_k$ the elements of the vector $\vec{f}$. 

 \begin{figure*}[ht!]
 	\includegraphics[width=.9\textwidth]{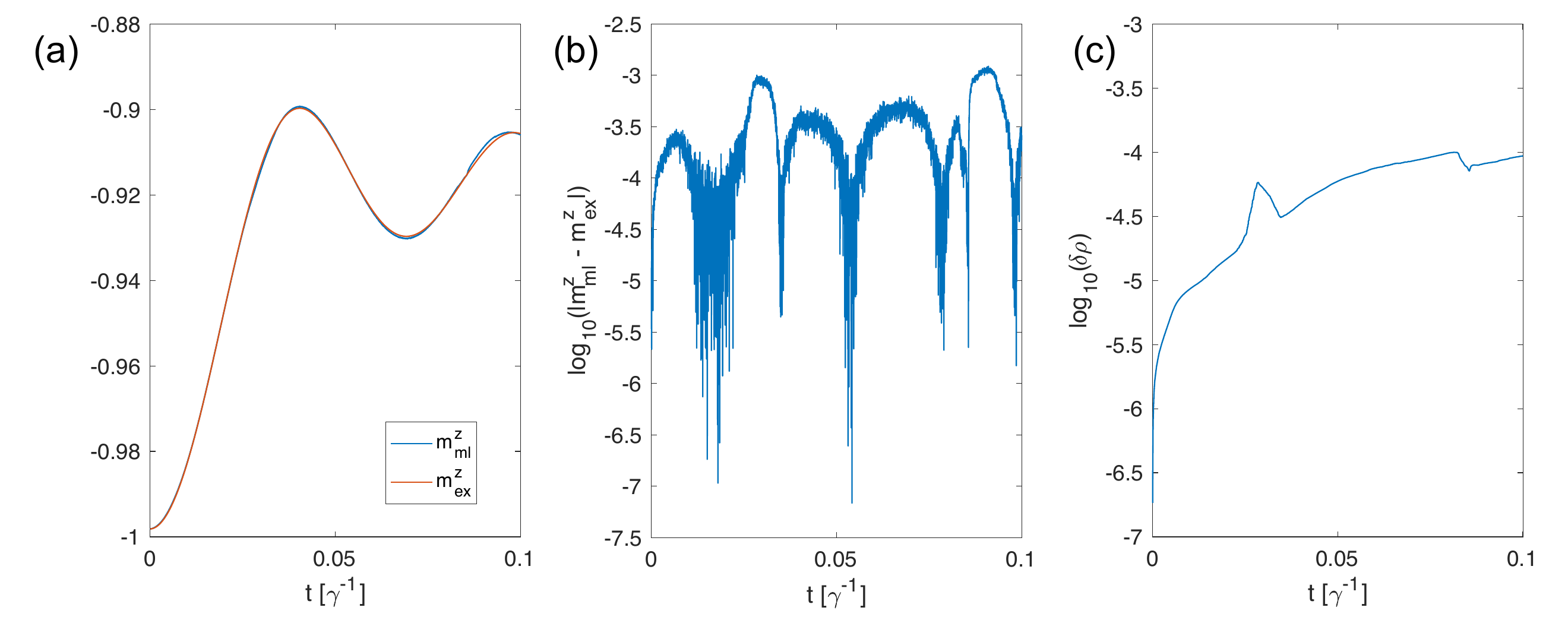}
 	\caption{Results for a chain of 5 spins with periodic boundary conditions and $B=10\gamma$, $J_x = 20 \gamma$, $J_y = 0$ and $J_z = 10\gamma$. $M = \tilde{M} = 20$, the sample size was $N_S = 10^6$ and the time step of 4-th order Runge-Kutta integration was $\delta t = 2 \times 10^{-5} \gamma^{-1}$. {\bf (a)} Magnetization for the neural network approximation, $\langle \sigma^z \rangle$ (blue) and the exact solution, $\langle \sigma^z \rangle_e$ (orange).  {\bf (b)} Log plot of difference between $\langle \sigma^z \rangle$ and $\langle \sigma^z \rangle_e$. {\bf (c)} Average deviation of the matrix element of $\rho$ from the exact density matrix $\rho_e$ as given by Eq. (\ref{eq:drho-def}).}
 	\label{fig:example1}
 \end{figure*} 

{\bf Stochastic sampling:} The expressions in Eqs. (\ref{eq:eom-var-1}-\ref{eq:eom-var-2}) cannot be exactly computed for systems with a large number of quantum particles. However, those quantum expectations can be conveniently interpreted as statistical expectation values over the probability distribution 
\begin{equation} \label{eq:distribution}
	p(\vec{l},\vec{r}) = |\rho_{\vec{l},\vec{r}}|^2, 
\end{equation}
in analogy to the concept in static and time-dependent variational Monte Carlo. 
The elements of $\mathcal{S}$ and $\vec{f}$ can thus also be written as,
\begin{equation}
	S_{k,p} \propto \mathrm{Re} \langle  O_k^\dag O_{p}  \rangle_{p}  \quad \text{and} \quad
	f_{k} \propto \mathrm{Re} \langle  O_k^\dag \mathcal{L}^{\mathrm{res}} \rangle_{p}, 
\end{equation}
%\begin{align}
%	S_{k,p} & \propto \mathrm{Re} \langle  O_k^\dag O_{p}  \rangle_{p}  \label{eq:exp-values-2} \\
%	f_{k} & \propto \mathrm{Re} \langle  O_k^\dag \mathcal{L}^{\mathrm{res}} \rangle_{p},  \label{eq:exp-values-3}
%\end{align}
where $\langle \dots \rangle_{p}$ denotes a statistical expectation value of the probability distribution $p$ as in Eq. (\ref{eq:distribution}), and we have introduced the following estimator for the Liouvillian, 
\begin{equation} \label{eq:local_liouvi}
	\mathcal{L}_{\vec{l_1},\vec{r_1};\vec{l_2},\vec{r_2}}^{\mathrm{res}} = \sum_{\vec{l_2},\vec{r_2}} \frac{\mathcal{L}_{\vec{l_1},\vec{r_1};\vec{l_2},\vec{r_2}} \rho_{\vec{l_2},\vec{r_2}}}{ \rho_{\vec{l_1},\vec{r_1}}}
\end{equation} 

In addition to having a stochastic strategy for solving the variational equations of motion, it is also important to provide an efficient scheme 
to compute expectation values of physical observables. Consider the expectation value of a generic observable $X$,  
\begin{equation}\label{eq:expect}
	\langle X \rangle = \text{Tr} \{ X \rho \} = \sum_{\vec{l},\vec{m}} X_{\vec{l},\vec{m}} \rho_{\vec{m},\vec{l}}.
\end{equation}
Estimates of $\langle X \rangle$ can be obtained in this case as statistical averages over the probability distribution $q(\vec{l}) = \rho_{\vec{l},\vec{l}}$, such that 

\begin{equation}
	\langle X \rangle \simeq \langle X^{\mathrm{loc}}\rangle_{q}, \quad \text{where} \quad 
	X^{\mathrm{loc}}_{\vec{l},\vec{l}} = \sum_{\vec{m}} \frac{X_{\vec{l},\vec{m}} \rho_{\vec{m},\vec{l}}}{\rho_{\vec{l},\vec{l}}}
\end{equation}
In all cases of physical relevance, observables $X$ have a sparse representation, and computing the estimator  $X^{\mathrm{loc}}_{\vec{l},\vec{l}} $ can be efficiently realized.
Notice that, while possible, sampling over $p(\vec{l},\vec{m})$ to compute physical expectation values would entail a much less efficient statistical estimator for  $\langle X \rangle$. This would further require to stochastically estimate the normalization factor, which is instead automatically taken into account when sampling from $q(\vec{l})$. 
%Furthermore, the Schwarz inequality guarantees that the summands do not diverge since $(\rho_{\vec{m},\vec{l}}/\rho_{\vec{l},\vec{l}}) \, (\rho_{\vec{l},\vec{m}}/\rho_{\vec{m},\vec{m}}) \leq 1$.
In this work we use two independent Markov-Chain Monte Carlo schemes to obtain samples both from  $p(\vec{l},\vec{r})$ and from $q(\vec{l})$ at each instant of time, as explained in the Supplementary Material.

{\bf Results:}
 To test the accuracy of our method, we consider  an anisotropic Heisenberg model with Hamiltonian
\begin{equation} \label{eq:model1}
	H = \sum_{j=1}^N B \sigma_j^z + \sum_{<j,l>} \sum_{a=x,y,z} J_a \sigma_j^a \sigma_l^a,
\end{equation} 
where $\sum_{<j,l>}$ denotes the sum over all nearest neighbors, and dissipator
\begin{equation} \label{eq:dissipator_model}
	\mathcal{D}[\rho] = \frac{\gamma}{2} \sum_{j=1}^N \left( 2\sigma_j^- \rho \sigma_j^+ - \sigma_j^+ \sigma_j^- \rho - \rho \sigma_j^+ \sigma_j^- \right) 
\end{equation}
%For computing the matrices $S$ and $f$ for the stochastic reconfiguration approach, we need to calculate the so called local Liouvillian, see Eq. (\ref{eq:exp-values-4}),
%The explicit forms of the terms for the model in Eqs. (\ref{eq:model1}) and (\ref{eq:dissipator_model}) are given in Supplemental Material.
Whereas our method can be applied equally to one and two dimensional lattices, we here present examples for one dimensional lattices where we can compare the results to matrix product state simulations.
We consider two applications. First we compare the time evolution of a density matrix as obtained from Eq. (\ref{eq:para_derivative_2}) to the exact time evolution of the density matrix for a small size model where the full master equation (\ref{eq:master}) can be numerically integrated. Then we show that our method correctly finds stationary states for a larger model, that can no longer be fully integrated but where a Matrix Product State (MPS) representation of $\rho$ \cite{hartmann_polariton_2010} allows to find the stationary state.

 To quantify the accuracy of a time evolution for $\rho$ as obtained from Eq. (\ref{eq:para_derivative_2}), we consider two quantities. {\bf (i)} The average deviation of the matrix element of $\rho$ from the exact density matrix $\rho^e$ iss given by 
 \begin{equation}\label{eq:drho-def}
 	\delta\rho = \frac{1}{2^{2N}} \sum_{j,l} |\rho_{j,l} - \rho_{j,l}^e|^2 = \frac{||\rho-\rho^e||_2^2}{2^{2N}}
 \end{equation}
 where $|| x ||$ is the 2-norm of a matrix $x$.  
 {\bf (ii)} To get an accuracy test in terms of physical observables we compare the magnetization for our approximation
 \begin{equation} \label{eq:magnetization}
 	m^z_{\text{ml}} = \frac{1}{N} \sum_j \langle \sigma_j^z \rangle
 \end{equation}
 to the magnetization for an exact solution $m^z_{\text{ex}} = \frac{1}{N} \sum_j \langle \sigma_j^z \rangle_{\text{exact}}$. 
 
  Results for a linear chain with $N=5$ spins and periodic boundary conditions are presented in Fig. \ref{fig:example1} and clearly show that the parameterization of the density matrix $\rho$ in terms of the neural network in Fig. \ref{fig:network} provides a very good approximation to the dissipative quantum dynamics of mixed states.

To show that our method correctly finds stationary states for models where the full density matrix can no longer be computed, we test whether $\mathcal{L}\rho = 0$. To this end we compute
\begin{equation} \label{eq:mag_liouv}
		\delta \mathcal{L} = \langle \left| \mathcal{L}^{\mathrm{res}} \right|\rangle_{p}.
\end{equation}
Since $\delta \mathcal{L} = \sum_{\vec{l_1},\vec{r_1}} |\rho_{\vec{l_1},\vec{r_1}}| \, | (\mathcal{L} \rho)_{\vec{l_1},\vec{r_1}}|$ this tests whether all matrix elements $(\mathcal{L} \rho)_{\vec{l_1},\vec{r_1}}$ vanish. Moreover, the measure $\delta\mathcal{L}$ weights the matrix elements of $\mathcal{L} \rho$ according to the relevance for the state $\rho$ and is very economic to compute. $\delta \mathcal{L}$ can thus be computed as a test for the convergence to the stationary state, even if the properties of the latter are completely unknown. Notice that since $\delta \mathcal{L}$ can be efficiently estimated (as well as other related quantities such as $|\mathcal{L}\rho|^2 \propto \langle \left| \mathcal{L}^{\mathrm{res}} \right|^2\rangle_{p}$, at the same cost of appliying $\mathcal{L}$ once, it is in principle possible to devise an alternative variational optimization scheme that directly minimizes $\delta \mathcal{L}$, if only the stationary state is of interest. 

\begin{figure}
 	\includegraphics[width=\columnwidth]{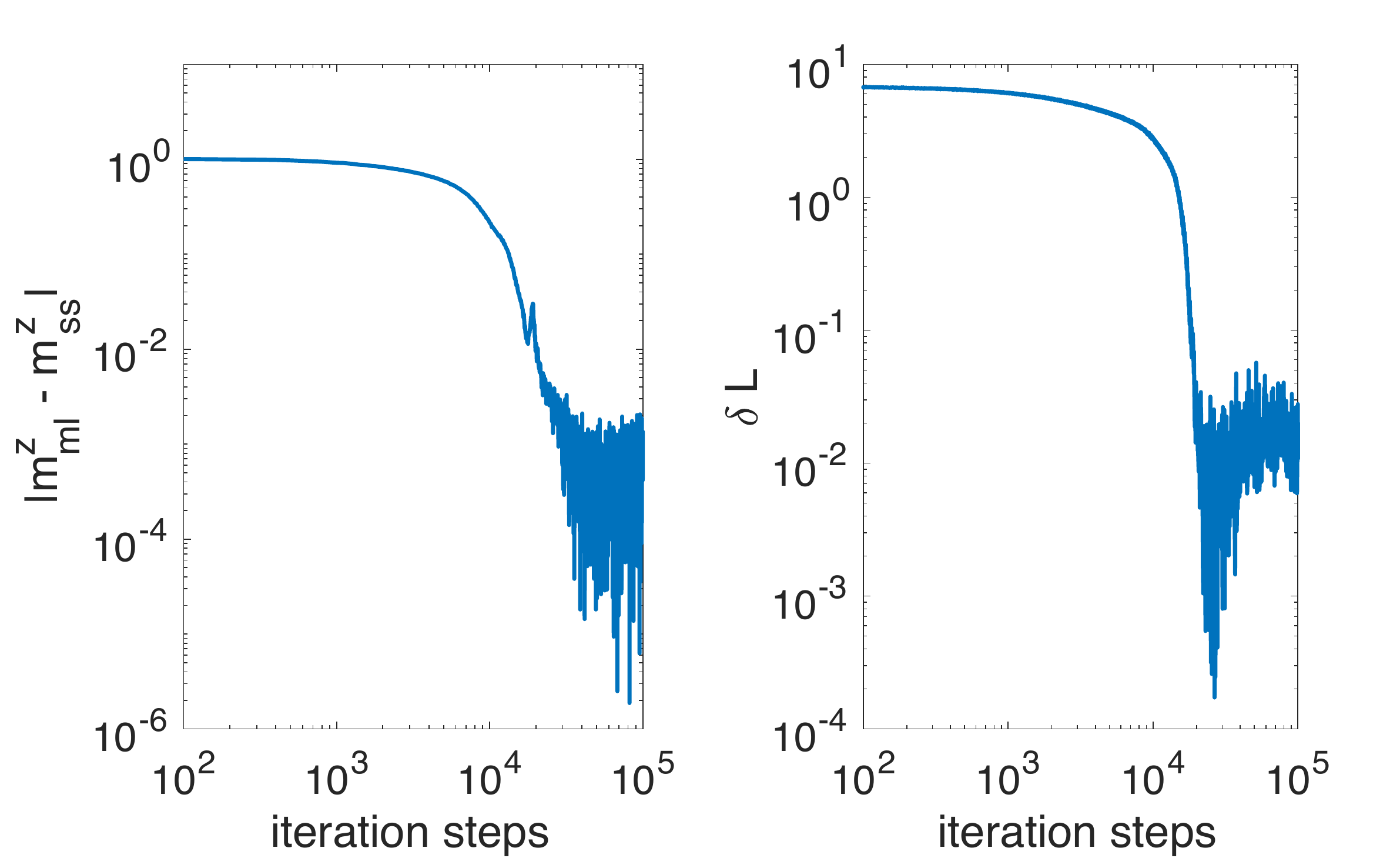}
 	\caption{Results for a chain of 16 spins with open boundary conditions $B=10\gamma$, $J_x = \gamma$, $J_y = 0$ and $J_z = 0$. $M = \tilde{M} = 12$, the sample size was $N_S = 2 \times 10^5$ and a 2nd order Runge-Kutta integration with adaptive step size was used. {\bf (a)} Difference between $m^z_{\text{ml}}$,  see Eq. (\ref{eq:magnetization}), and $m^z_{\text{ss}} = -15.9286$ as found from an integration with MPS. and {\bf (b)} magnitude of $\mathcal{L}\rho$ as quantified by $\delta L$ given in Eq. (\ref{eq:mag_liouv})}
 	\label{fig:example2}
 \end{figure}

In addition to computing $\delta\mathcal{L}$, we also test whether the magnetization, see Eq. (\ref{eq:magnetization}) approaches the steady state magnetization $m^z_{\text{ss}}= \lim_{t \to \infty} \frac{1}{N} \sum_j \langle \sigma_j^z \rangle$, which we obtain from an integration with a MPS representation of the density matrix $\rho$.
Results for the approach to the stationary state of a chain with $N=16$ spins and open boundary conditions are presented in Fig. \ref{fig:example2} and show that the stationary state is found with high accuracy. 
 For finding stationary states, we make use of the fact that the parameterization (\ref{eq:parameter-torlai-2}) always guarantees a physically valid state. If we do not require to correctly model the dynamics for all times, we can thus choose the integration time step larger and still find convergence to the correct stationary state.
 
The example presented here, features a moderately correlated state where an MPS representation of the steady state with bond dimension $D=17$ suffices to compute $m^z_{\text{ss}}$ with an accuracy of $10^{-4}$. Hence, the MPS representation here requires $4*[(N-2) \times D^{2} + 2*D] = 16320$ parameters (the pre-factor 4 is the physical dimension for mixed states of spin systems), whereas the neural network representation achieves a comparable approximation with only $(N+1) (M + \tilde{M}) + N = 424$ parameters. Scenarios with stronger spin-spin interactions would require more variational parameters and larger sample sizes, increasing the numerical effort of the method. 

{\bf Conclusions:}
We have introduced a neural-network based approach to numerically modeling the quantum dynamics and stationary quantum states of open or dissipative quantum many-body systems. Our results show that both, the dynamics and stationary states of such systems can be obtained with high accuracy. In this work we have shown one-dimensional systems, in order to provide benchmarks with existing approaches. Several extensions of our approach can be envisaged for future research. From the point of view of applications, the study of two-dimensional lattices does not present conceptual difficulties, and will represent an interesting opportunity for our method. From the methodological point of view, schemes targeting only the stationary state can also be efficiently implemented, using the same ideas introduced to compute $\delta \mathcal{L}$ in this work. 

{\bf Acknowledgements:} 
MJH thanks Heriot-Watt University for support. We acknowledge stimulating discussions with V. Savona, C. Ciuti, F. Vicentini, and G. Torlai. 

{\bf Note added:}
Shortly after completion of this work, references \cite{yoshioka_constructing_2019,vicentini_variational_2019,nagy_variational_2019} appeared in preprint, 
which discuss similar strategies to study stationary states of open quantum many-body systems by using complex-valued neural-networks.

\bibliography{biblio}

\pagebreak

\appendix

\begin{widetext}
\begin{center}
	{\bf SUPPLEMENTAL MATERIAL}
\end{center}

\section{Proof of Minimum}\label{sec:minimum}

$\mathcal{S}$ and $\vec{f}$ can be written as,
\begin{align}
	S_{k,p} & = \vec{\rho}^\dag O_k^\dag O_p \vec{\rho}  + \vec{\rho}^\dag O_p^\dag O_k \vec{\rho} = \frac{\partial \vec{\rho}^\dag}{\partial \alpha_k} \frac{\partial \vec{\rho}}{\partial \alpha_p} + \frac{\partial \vec{\rho}^\dag}{\partial \alpha_p} \frac{\partial \vec{\rho}}{\partial \alpha_k} \nn \\
	f_k & = \vec{\rho}^\dag O_k^\dag \mathcal{L} \vec{\rho} + \vec{\rho}^\dag \mathcal{L}^\dag O_k \vec{\rho} = \frac{\partial \vec{\rho}^\dag}{\partial \alpha_k} \mathcal{L} \vec{\rho} + \vec{\rho}^\dag \mathcal{L}^\dag \frac{\partial \vec{\rho}}{\partial \alpha_k}\nn
\end{align}
The $\dot{\alpha}_k$ are real and we have for any vector with real elements $v_k$, 
\begin{align}
	\sum_{k,p} v_k S_{k,p} v_p = & \sum_{k,p} \left( v_k \frac{\partial \vec{\rho}^\dag}{\partial \alpha_k} \frac{\partial \vec{\rho}}{\partial \alpha_p} v_p + v_p \frac{\partial \vec{\rho}^\dag}{\partial \alpha_p} \frac{\partial \vec{\rho}}{\partial \alpha_k} v_k \right) = 2 \vec{\nu}^\dag \vec{\nu} \geq 0
\end{align}
where $\vec{\nu} = \sum_k \frac{\partial \vec{\rho}}{\partial \alpha_k} v_k$. Therefore, the matrix $S$ is positive semidefinite and the solution to Eq. (\ref{eq:para_derivative}) is indeed the minimum of $\delta$.

\section{Fubini-Study norm}
Alternatively to the 2-norm, the Fubini-Study norm can be employed to derive the approximation. For this approach one minimizes
\begin{equation}
	\gamma(\vec{\sigma},\vec{\mu}) = \arccos \sqrt{\frac{|\vec{\sigma}^\dag \cdot \vec{\mu}|^2}{(\vec{\sigma}^\dag \cdot \vec{\sigma}) (\vec{\mu}^\dag \cdot \vec{\mu})}}
\end{equation}
for $\sigma = \sum_k \dot{\alpha}_k O_k \vec{\rho}$ and $\mu = \mathcal{L} \vec{\rho}$. This leads to the same system of ODEs as given in Eq. (7) of the main text, but where $S$ and $f$ read,
\begin{align}
	S_{k,k'} & = \vec{\rho}^\dag O_{k}^\dag O_{k'}\vec{\rho} + \vec{\rho}^\dag O_{k'}^\dag O_{k}\vec{\rho} 
	- \vec{\rho}^\dag O_{k}^\dag \vec{\rho} \,  \vec{\rho}^\dag O_{k'} \vec{\rho} - \vec{\rho}^\dag O_{k'}^\dag \vec{\rho} \,  \vec{\rho}^\dag O_{k} \vec{\rho} \\
	f_k & = \vec{\rho}^\dag O_{k}^\dag \mathcal{L} \vec{\rho} + \vec{\rho}^\dag O_{k} \mathcal{L}^\dag \vec{\rho} - 
	\vec{\rho}^\dag O_{k}^\dag \vec{\rho} \,  \vec{\rho}^\dag \mathcal{L} \vec{\rho} - 
	\vec{\rho}^\dag O_{k} \vec{\rho} \,  \vec{\rho}^\dag \mathcal{L}^\dag \vec{\rho}
\end{align}

%The expectation values in $\mathcal{S}$ and $\vec{f}$ can be written as,
%\begin{align}
%	\vec{\rho}^\dag O_k \vec{\rho} & = \sum_{\vec{l},\vec{r}} |\rho_{\vec{l},\vec{r}}|^2 \frac{\partial \ln ( \rho_{\vec{l},\vec{r}}) }{\partial \alpha_k} \\
%	\vec{\rho}^\dag O_k O_{k'} \vec{\rho} & = \sum_{\vec{l},\vec{r}} |\rho_{\vec{l},\vec{r}}|^2 \frac{\partial \ln ( \rho_{\vec{l},\vec{r}})}{\partial \alpha_k} \, \frac{\partial \ln ( \rho_{\vec{l},\vec{r}})}{\partial \alpha_{k'}} \nn \\
%	\vec{\rho}^\dag \mathcal{L} \vec{\rho} & = \sum_{\vec{l_1},\vec{r_1}} \sum_{\vec{l_2},\vec{r_2}} \rho_{\vec{l_1},\vec{r_1}}^\star \mathcal{L}_{\vec{l_1},\vec{r_1};\vec{l_2},\vec{r_2}} \rho_{\vec{l_2},\vec{r_2}} \nn \\
%			& = \sum_{\vec{l_1},\vec{r_1}} \sum_{\vec{l_2},\vec{r_2}} |\rho_{\vec{l_1},\vec{r_1}}|^2 \frac{\mathcal{L}_{\vec{l_1},\vec{r_1};\vec{l_2},\vec{r_2}} \rho_{\vec{l_2},\vec{r_2}}}{ \rho_{\vec{l_1},\vec{r_1}}}\nn \\
%	\vec{\rho}^\dag O_k \mathcal{L} \vec{\rho} & = \sum_{\vec{l_1},\vec{r_1}} \sum_{\vec{l_2},\vec{r_2}} \rho_{\vec{l_1},\vec{r_1}}^\star \frac{\partial \ln ( \rho_{\vec{l_1},\vec{r_1}})}{\partial \alpha_k} \, \mathcal{L}_{\vec{l_1},\vec{r_1};\vec{l_2},\vec{r_2}} \rho_{\vec{l_2},\vec{r_2}} \nn \\
%& = \sum_{\vec{l_1},\vec{r_1}} \sum_{\vec{l_2},\vec{r_2}} |\rho_{\vec{l_1},\vec{r_1}}|^2 \frac{\partial \ln ( \rho_{\vec{l_1},\vec{r_1}})}{\partial \alpha_k} \, \frac{\mathcal{L}_{\vec{l_1},\vec{r_1};\vec{l_2},\vec{r_2}} \rho_{\vec{l_2},\vec{r_2}}}{ \rho_{\vec{l_1},\vec{r_1}}}	\nn
%\end{align}

\section{Logarithmic derivatives} \label{sec:log-devs}
The logarithmic derivatives read,
\begin{align} 
\frac{\partial\ln(\rho_{\vec{l},\vec{r}})}{\partial \text{Re}[a_{n}]} & = l_{n}+r_{n} \label{eq:logdevs-Torlai-1}\\
\frac{\partial\ln(\rho_{\vec{l},\vec{r}})}{\partial \text{Im}[a_{n}]} & = i(l_{n}-r_{n})\\
\frac{\partial\ln(\rho_{\vec{l},\vec{r}})}{\partial \text{Re}[b_{k}]} & =
\xi_{b,W}(k,\vec{l}) + [\xi_{b,W}(k,\vec{r})]^*\\
\frac{\partial\ln(\rho_{\vec{l},\vec{r}})}{\partial \text{Im}[b_{k}]} & =
i \xi_{b,W}(k,\vec{l}) - i [\xi_{b,W}(k,\vec{r})]^*\\
\frac{\partial\ln(\rho_{\vec{l},\vec{r}})}{\partial \text{Re}[W_{k,n}]} & =
l_n \xi_{b,W}(k,\vec{l}) + r_n [\xi_{b,W}(k,\vec{r})]^*\\
\frac{\partial\ln(\rho_{\vec{l},\vec{r}})}{\partial \text{Im}[W_{k,n}]} & =
i l_n \xi_{b,W}(k,\vec{l}) - i r_n [\xi_{b,W}(k,\vec{r})]^*\\
\frac{\partial\ln(\rho_{\vec{l},\vec{r}})}{\partial \text{Re}[c_{p}]} & =
2 \, \xi_{c,U}(p,\vec{l},\vec{r}) \\
\frac{\partial\ln(\rho_{\vec{l},\vec{r}})}{\partial \text{Re}[U_{p,n}]} & =
(l_n + r_n) \, \xi_{c,U}(p,\vec{l},\vec{r}) \\
\frac{\partial\ln(\rho_{\vec{l},\vec{r}})}{\partial \text{Im}[U_{p,n}]} & =
i (l_n - r_n) \, \xi_{c,U}(p,\vec{l},\vec{r}) \label{eq:logdevs-Torlai-2}
\end{align}
where
%\begin{align}
%	\xi_{b,W}(k,\vec{n}) & = \frac{\exp\left[b_k + \sum_{j=1}^N W_{k,j} n_j\right]}{1 + \exp\left[b_k + \sum_{j=1}^N W_{k,j} n_j\right]}\\
%	\xi_{c,U}(p,\vec{l},\vec{r}) & = \frac{\exp\left[c_p + c_p^* + \sum_{j=1}^N U_{p,j} l_j + \sum_{j=1}^N U_{p,j}^* r_j\right]}{1 + \exp\left[c_p + c_p^* + \sum_{j=1}^N U_{p,j} l_j + \sum_{j=1}^N U_{p,j}^* r_j\right]}
%\end{align}
%For the alternative representation in Eq. (\ref{eq:parameter-torlai-2}), we get the same derivatives as in Eqs. (\ref{eq:logdevs-Torlai-1}) - (\ref{eq:logdevs-Torlai-2}), but now with 
\begin{align}
	\xi_{b,W}(k,\vec{n}) & = \tanh\left[b_k + \sum_{j=1}^N W_{k,j} n_j\right] \\
	\xi_{c,U}(p,\vec{l},\vec{r}) & = \tanh\left[c_p + c_p^* + \sum_{j=1}^N U_{p,j} l_j + \sum_{j=1}^N U_{p,j}^* r_j\right]
\end{align}
Since they all involve sums of exponentially many terms, we will compute all expectation values stochastically via Monte Carlo sampling.

\subsection{Sampling for expectation values}
\label{sec:expect-vals-app}
We here provide the arguments why one needs to sample from the distribution formed by the diagonal elements of the density matrix for computing expectation values of observables. Sampling over $p(\vec{l},\vec{m})$ would not work for computing such expectation values since this would require finding the normalization via
\begin{equation}
	\mathcal{N} = \sum_{\vec{l}} \rho_{\vec{l},\vec{l}} = \sum_{\vec{l},\vec{m}} p(\vec{l},\vec{m}) \frac{\rho_{\vec{m},\vec{m}}}{|\rho_{\vec{l},\vec{m}}|^2} \approx \frac{1}{N_S} \sum_{\vec{l},\vec{r} \in \mathcal{S}} \frac{\rho_{\vec{m},\vec{m}}}{|\rho_{\vec{l},\vec{m}}|^2}
\end{equation} 
which may be flawed by the statistical behavior. Indeed if we look at the variance,
\begin{align}
	\Delta\mathcal{N}^2 & = \sum_{\vec{l},\vec{m}} p(\vec{l},\vec{m}) \left(\frac{\rho_{\vec{m},\vec{m}}}{|\rho_{\vec{l},\vec{m}}|^2}\right)^2 - \mathcal{N}^2 = \sum_{\vec{l},\vec{m}} \frac{\rho_{\vec{m},\vec{m}}^2}{|\rho_{\vec{l},\vec{m}}|^2} - \mathcal{N}^2 
\end{align} 
we find that it could diverge for $|\rho_{\vec{l},\vec{m}}| \to 0$. This can become problematic due to the Schwarz inequality $|\rho_{\vec{l},\vec{m}}|^2 \leq \rho_{\vec{l},\vec{l}} \rho_{\vec{m},\vec{m}}$.

\section{Local estimator of the Liouvillian} \label{sec:local_liouvi}
For computing the matrices $S$ and $f$ for the stochastic reconfiguration approach, we need to calculate the so called local Liouvillian, see Eq. (10) of the main text. We first consider the term $-i [H,\rho]$, for which the local Liouvillian reads,
\begin{equation*}
    \frac{-i\langle \vec{l}|[H,\rho]|\vec{r}\rangle}{\rho_{\vec{l},\vec{r}}} =-i\sum_{\vec{m}} H_{\vec{l},\vec{m}}\frac{\rho_{\vec{m},\vec{r}}}{\rho_{\vec{l},\vec{r}}}+ i\sum_{\vec{m}}\frac{\rho_{\vec{l},\vec{m}}}{\rho_{\vec{l},\vec{r}}} H_{\vec{m},\vec{r}}	
\end{equation*}
%
%\begin{multline}
%\frac{-i\langle l|[H,\rho]|r\rangle}{\rho(l,r)}=i\sum_{r^{\prime}}\frac{\rho(l,r^{\prime})}{\rho(l,r)}H_{rr^{\prime}}^{\star}+\\
%-i\sum_{l^{\prime}}\frac{\rho(l^{\prime},r)}{\rho(l,r)}H_{ll^{\prime}},
%\end{multline}
which can be computed efficiently for k-local Hamiltonians, with the
same complexity of the local energy in standard variational calculations
for the ground-state. 
For the Hamiltonian in Eq. (\ref{eq:model1}), we have in one dimension
\begin{align*}
	H_{\vec{m},\vec{n}} & = \sum_{j=1}^N \langle \vec{m}|H_j|\vec{n} \rangle\\
\langle \vec{m}|H_j|\vec{n} \rangle & = B \langle \vec{m}|\sigma_j^z |\vec{n}\rangle + J_x \langle \vec{m}|\sigma_j^x \sigma_{j+1}^x|\vec{n}\rangle + J_y \langle \vec{m}|\sigma_j^y \sigma_{j+1}^y|\vec{n}\rangle + J_z \langle \vec{m}|\sigma_j^z \sigma_{j+1}^z|\vec{n}\rangle
\end{align*}
For the dissipator, $\mathcal{D}[\rho]$, we get,
\begin{align*}
    \mathcal{D}[\rho]&=\sum_{j=1}^{N}\mathcal{D}_{j}[\rho]\\
	\frac{\langle \vec{l}|\mathcal{D}_{j}[\rho]|\vec{r} \rangle}{\rho_{\vec{l},\vec{r}}}& = \gamma \sum_{\vec{l}^{\prime}, \vec{r}^{\prime}} \langle \vec{l}|\sigma_{j}^-|\vec{l}^{\prime}\rangle\langle \vec{r}^\prime |\sigma_{j}^+|\vec{r}\rangle \frac{\rho_{\vec{l}^\prime,\vec{r}^\prime}}{\rho_{\vec{l},\vec{r}}} -\frac{\gamma}{2}\sum_{\vec{m}}\langle \vec{l}| \sigma_j^+ \sigma_j^- |\vec{m}\rangle \frac{\rho_{\vec{m},\vec{r}}}{\rho_{\vec{l},\vec{r}}} -\frac{\gamma}{2}\sum_{\vec{m}}\langle \vec{m}|\sigma_j^+ \sigma_j^- |\vec{r}\rangle \frac{\rho_{\vec{l},\vec{m}}}{\rho_{\vec{l},\vec{r}}}.\nn
\end{align*}
We use here a notation, where $n_j = \pm 1$. Thus we get,
\begin{align*}
	\langle \vec{m}|\sigma_j^z |\vec{n}\rangle & = n_j \, \delta_{\vec{m},\vec{n}} \nn \\
	\langle \vec{m}|\sigma_j^x \sigma_{j+1}^x|\vec{n}\rangle & = \delta_{n_j, -m_j} \, \delta_{n_{j+1}, -m_{j+1}} \, \prod_{i \neq j,j+1} \delta_{m_i,n_i} \nn \\
	\langle \vec{m}|\sigma_j^y \sigma_{j+1}^y|\vec{n}\rangle & = \delta_{n_j, -m_j} \, \delta_{n_{j+1}, -m_{j+1}} (\delta_{n_j,-n_{j+1}} - \delta_{n_j, n_{j+1}}) \prod_{i \neq j,j+1} \delta_{m_i,n_i} \nn \\
	\langle \vec{m}|\sigma_j^z \sigma_{j+1}^z|\vec{n}\rangle & = n_j n_{j+1} \, \delta_{\vec{m},\vec{n}} \nn \\
	\langle \vec{m}|\sigma_j^+ \sigma_j^- |\vec{n}\rangle & = \delta_{n_j,1} \, \delta_{\vec{m},\vec{n}} \nn \\
	\langle \vec{m}|\sigma_{j}^-|\vec{n}\rangle & = \delta_{n_j,1} \delta_{m_j,-1} \prod_{i \neq j} \delta_{m_i,n_i}\nn \\
	\langle \vec{m}|\sigma_{j}^+|\vec{n}\rangle & = \delta_{n_j,-1} \delta_{m_j,1} \prod_{i \neq j} \delta_{m_i,n_i}
\end{align*}
where the $\delta_{j,l}$ are Kronecker deltas. Using the notation $\rho (l_1,\dots,l_n;r_1,\dots, r_n) =\rho_{l_1,\dots,l_n;r_1,\dots, r_n}$ with $n=2^N$, we get,
\begin{align*}
	\frac{\langle \vec{l}|\mathcal{D}_{j}[\rho]|\vec{r} \rangle}{\rho_{\vec{l},\vec{r}}}& = \gamma \delta_{l_j,-1} \delta_{r_j,-1} \frac{\rho (\dots,l_{j-1},1,l_{j+1}, \dots;\dots,r_{j-1},1,r_{j+1}, \dots )}{\rho (\dots,l_{j-1},-1,l_{j+1}, \dots;\dots,r_{j-1},-1,r_{j+1}, \dots )} -\frac{\gamma}{2}( \delta_{l_j,1} + \delta_{r_j,1}) 
\end{align*}
and
\begin{align*}
	\frac{-i\langle \vec{l}|[H_j,\rho]|\vec{r}\rangle}{\rho_{\vec{l},\vec{r}}} & = - i B (l_j-r_j) -i J_z (l_j l_{j+1} - r_j r_{j+1}) \\
	& -i J_x \left[\frac{   \rho (\dots,-l_j,-l_{j+1},\dots; \dots )}{ \rho (\dots ,l_j,l_{j+1}, \dots;\dots )} - \frac{ \rho (\dots ;\dots,-r_j,-r_{j+1}, \dots )}{ \rho (\dots ;\dots,r_{j},r_{j+1}, \dots )}\right]\nn\\
	& -i J_y \left[(\delta_{l_j,-l_{j+1}} - \delta_{l_j,l_{j+1}}) \frac{ \rho (\dots ,-l_j,-l_{j+1},\dots;\dots )}{ \rho (\dots ,l_j,l_{j+1}, \dots;\dots )} - (\delta_{r_j,-r_{j+1}} - \delta_{r_j, r_{j+1}})\frac{ \rho (\dots ;\dots,-r_j,-r_{j+1}, \dots )}{ \rho (\dots ;\dots,r_{j},r_{j+1}, \dots )}\right]\nn
\end{align*}

\section{Moves of the Metropolis Sampling}

For sampling from the distribution $p(\vec{l},\vec{r})$,  we considered four types of moves,
\begin{enumerate}
	\item one index, either left or right ($l_j$ or $r_j$) is flipped.
	\item the left and right indices $l_j$ and $r_j$, corresponding to one site $j$ are both flipped.
	\item neighboring left indices $l_j$ and $l_{j+1}$ or right indices $r_j$ and $r_{j+1}$ are flipped.
	\item a new configuration is drawn from a uniform distrbution.
\end{enumerate}
Whereas the moves 1-3 occur with the same probability, the likelihood for move 4 was chosen to be 100 times smaller.

In turn for the sampling from the distribution $q(\vec{l},\vec{l})$, we considered three types of moves,
\begin{enumerate}
	\item one index $l_j$ is flipped.
	\item neighboring indices $l_j$ and $l_{j+1}$ are flipped.
	\item a new configuration is drawn from a uniform distrbution.
\end{enumerate}
Here again, the last move was chosen to occur 100 times less often than the other two.

\end{widetext}
\end{document}